\documentclass
[preprint,superscriptaddress,pre,a4paper,twocolumn,showpacs,nobibnotes,nofootinbib,floats,balancelastpage,twoside,10pt]{revtex4}%
\usepackage{amsfonts}
\usepackage{amsmath}
\usepackage{amssymb}
\usepackage{graphicx}%
\begin{document}
\preprint{ }
\title{Temporal dynamics in the one-dimensional quantum Zakharov equations for plasmas}
\author{A. P. Misra }
\email{apmisra@visva-bharati.ac.in}
\altaffiliation{Permanent address: Department of Mathematics, Visva-Bharati University,
Santiniketan-731 235, India.}

\affiliation{Department of \ Physics, Ume\aa \ University, SE-901 87 Ume\aa , Sweden.}
\author{S. Banerjee}
\email{santo.banerjee@polito.it}
\affiliation{Department of Mathematics, Politecnico di Torino, Turin, Italy.}
\affiliation{Micro and Nanotechnology Unit, Techfab s.r.l., Chivasso, Italy.}
\author{F. Haas }
\email{ferhaas@unisinos.br}
\affiliation{Department of Physics, Ume\aa \ University, SE-901 87 Ume\aa , Sweden.}
\author{P. K. Shukla}
\email{ps@tp4.rub.de.}
\affiliation{Institut f\"{u}r Theoretische Physik IV, Ruhr-Universit\"{a}t Bochum, D-44780
Bochum, Germany.}
\author{L. P. G. Assis}
\email{lpgassis@ufrrj.br}
\affiliation{Grupo de F\'{\i}sica Te\'orica e Matem\'atica F\'{\i}sica, Universidade
Federal Rural do Rio de Janeiro BR 465-07, 23851-180, Serop\'edica, Rio de
Janeiro, Brazil.}

\begin{abstract}
The temporal dynamics of the quantum Zakharov equations\ (QZEs) in one spatial
dimension, which describes the nonlinear interaction of quantum Langmuir waves
(QLWs) and quantum ion-acoustic waves (QIAWs) is revisited by considering
their solution as a superposition of three interacting wave modes in Fourier
space. Previous results in the literature are modified and rectified.
\ Periodic, chaotic as well as hyperchaotic behaviors of the Fourier-mode
amplitudes are identified by the analysis of Lyapunov exponent spectra and the
power spectrum. The periodic route to chaos is explained through an
one-parameter bifurcation analysis. The system is shown to be destabilized via
a supercritical Hopf-bifurcation. The adiabatic limits of the fully
spatio-temporal and reduced systems are compared from the viewpoint of
integrability properties.
\end{abstract}
\received{2 Dec., 2009}
\revised{15 Feb., 2010}
\accepted{16 Feb., 2010}
\startpage{1}
\endpage{102}
\maketitle
\section{Introduction}

The Zakharov equations (ZEs) are one of the most important models in plasma
physics community \cite{Zakharov}, in which high-frequency Langmuir waves
(LWs) are nonlinearly coupled with the low-frequency ion-acoustic waves
(IAWs). In this context, the quantum Zakharov equations (QZEs) are the
modified version of the classical ZEs, including a quantum correction
associated with the Bohm potential \cite{Garcia}. Such QZEs are also deduced
from a multiple time-scale technique applied to a set of quantum hydrodynamic
(QHD) equations under the quasineutrality assumption. Recent works (see e.g.
Refs. \cite{Marklund,Haas1,Misra1,Misra2}) indicate that much attention has
been paid to investigate the dynamics of such one-dimensional (1D) QZEs. Very
recently, a more comprehensive work on the dynamics of LWs has been studied by
Haas and Shukla in a three-dimensional quantum Zakharov system \cite{Haas2}.
The arrest of Langmuir wave collapse by quantum effects, predicted by a
variational approach in Ref. \cite{Haas2}, was later confirmed with rigorous
estimates and systematic asymptotic expansions by Simpson et al \cite{Simpson}.

In contrast to classical ZEs, the QZEs can not be reduced, in the adiabatic
limit, to a nonlinear Schr\"{o}dinger equation (NLSE) \cite{Garcia}. Rather,
it follows a coupled system for the envelope electric field and the density
fluctuation, whose complete integrability is not assured. However, the system
can be decoupled in the adiabatic as well as semiclassical limit
\cite{Garcia,Misra2}, where the solitons can be found to be more stable than
in fully degenerate cases \cite{Misra2}. \ Other recent developments on the
QZEs can be found in the literature
\cite{Marklund,Haas1,Misra1,Misra2,Haas2,Tang,Abdou,Wakil}. Marklund
\cite{Marklund} studied the kinetic theory of LWs interacting with quantum
ion-acoustic waves (QIAWs), where \ it was shown that the combined effects of
partial coherence and quantum correction tend to enhance the modulational
instability (MI) growth rate \cite{Marklund}. The temporal dynamics of the
QZEs has been studied by Haas by means of a variational approach \cite{Haas1}.
It has been found that the quantum coupling parameter plays a destabilizing
role on localized structures, or Langmuir wave packets. More mathematical
treatments on the QZEs are the analysis of the underlying Lie symmetry group
\cite{Tang} and the derivation of some exact solutions \cite{Abdou,Wakil}. A
Galerkin-type approximation was used by Misra et al. \cite{Misra1} to reduce
the QZEs to a set of ordinary differential equations (ODEs) for the temporal
dynamics. This system was shown to exhibit hyperchaos (more than one positive
Lyapunov exponent). \ However, while qualitatively correct, unfortunately the
reduced model considered in this work was flawed by some algebraic
inconsistencies.

The primary goal of the present work is to revisit the temporal behavior of
the QZEs as a superposition of three interacting wave modes in Fourier space
and to rectify the previous results \cite{Misra1}. In addition, the evidence
for the existence of periodic limit cycles, chaotic as well as hyperchaotic
attractors of the Fourier-mode amplitudes are presented through the analysis
of bifurcation diagram, power spectra as well\ as the Lyapunov exponents.
There are relevant differences of the respective periodic, chaotic and
hyperchaotic regimes in parameter space. Our findings thus extend both
qua\-li\-ta\-ti\-ve\-ly and quantitatively the previous results as well as
exhibit some new features not reported in the earlier investigations
\cite{Haas1,Misra1}.

\section{Simplified model}

The 1D QZEs read \cite{Garcia}
\begin{align}
&  i{\frac{\partial E}{\partial t}}+{\frac{\partial^{2}E}{\partial x^{2}}%
}-H^{2}{\frac{\partial^{4}E}{\partial x^{4}}}=n\,E\,,\label{e1}\\
&  {\frac{\partial^{2}n}{\partial t^{2}}}-{\frac{\partial^{2}n}{\partial
x^{2}}}+H^{2}{\frac{\partial^{4}n}{\partial x^{4}}}={\frac{\partial^{2}%
|E|^{2}}{\partial x^{2}}}\,, \label{e2}%
\end{align}
where $E=E(x,t)$ is the envelope electric field and $n=n(x,t)$ is the plasma
density fluctuation (measured from its equilibrium value). In Eqs.
(\ref{e1}--\ref{e2}), the same set of dimensionless quantities of Ref.
\cite{Garcia} is employed. In particular, $H=\hbar\,\omega_{i}/\kappa
_{B}\,T_{e}$ is a quantum coupling parameter expressing the ratio between the
ion plasmon energy and the electron thermal energy, where $\hbar$ is Planck's
constant divided by $2\pi$, $\omega_{i}$ is the ion plasma frequency,
$\kappa_{B}$ is the Boltzmann constant and $T_{e}$ is the electron fluid
temperature. The formal classical limit $H\rightarrow0$ yields the original
Zakharov system \cite{Zakharov}.

The finite-dimensional temporal dynamics of the QZEs was studied in Ref.
\cite{Misra1}. In this investigation, the appearance of hyperchaos was
established by means of the derivation of two positive Lyapunov exponents and
the analysis of the Kaplan-Yorke dimension. The dynamics of this
(low-dimensional) simplified system was con\-si\-de\-red for a wide parameter
range of the system, in\-clu\-ding the quantum coupling parameter $H$.
However, in the derivation of this simplified model the change of variables in
Eqs. (7) and (8) of this work is not fully consistent. Indeed, Eq. (3b) of
this paper implies $\dot{a}=-4\,n_{1}\,\sin\phi$, while Eq. (3c) would imply
$\dot{a}=-2\,n_{1}\,\sin\phi$, using the notations of Ref. \cite{Misra1}. The
reason for the contradiction is that the transformations [Eqs. (7), (8) of
\cite{Misra1}] do not respect the conservation of plasmon number as given by
Eq. (5) of the cited paper. Similar difficulties also appear in the literature
about the classical Zakharov system \cite{Sharma, Batra}, in the derivation of
reduced ordinary differential equations simulating the classical ($H=0$)
temporal as well as spatio-temporal system. Hence, in this Section we will
rederive the basic set of equations assuming a truncated expansion scheme and
produce the correct simplified model. The resulting system is then shown
nu\-me\-ri\-cally to be compatible with the existence of chaotic as well as
hyperchaotic attractors, in qualitative agreement with the conclusions in Ref.
\cite{Misra1}. \begin{figure}[ptb]
\begin{center}
\includegraphics[height=2.5in,width=2.5in]{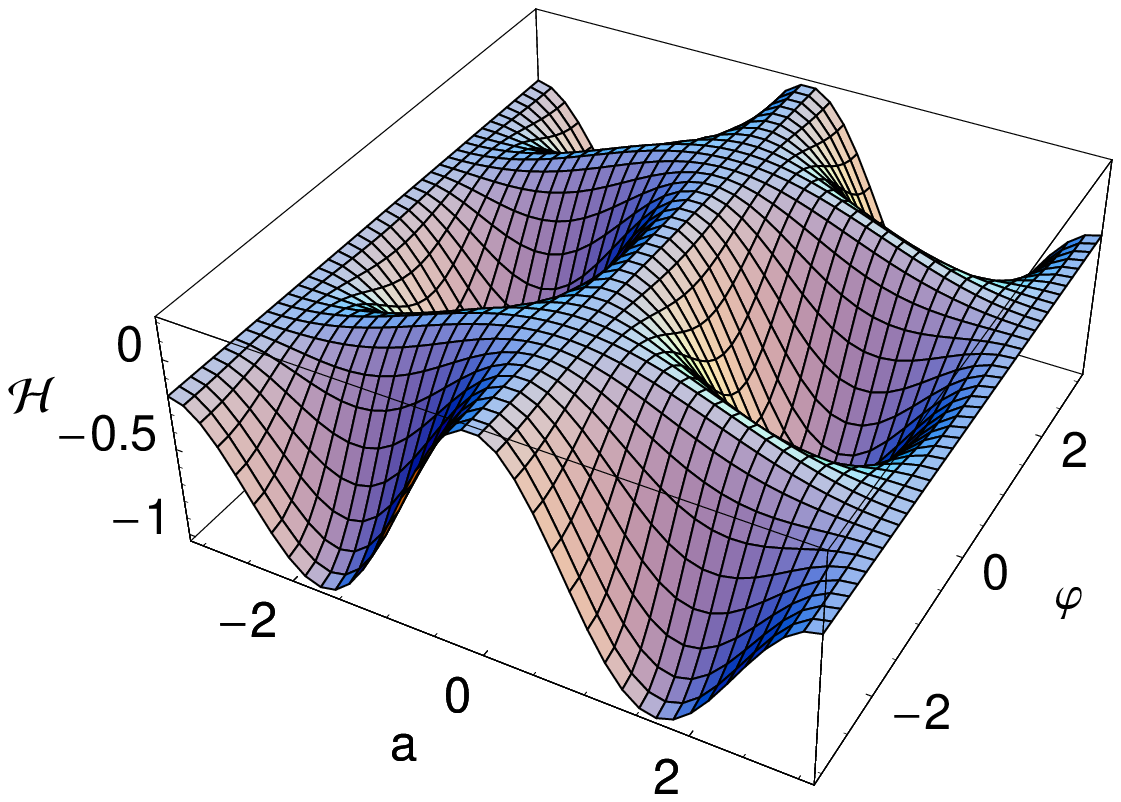}\newline%
\includegraphics[height=2.5in,width=2.5in]{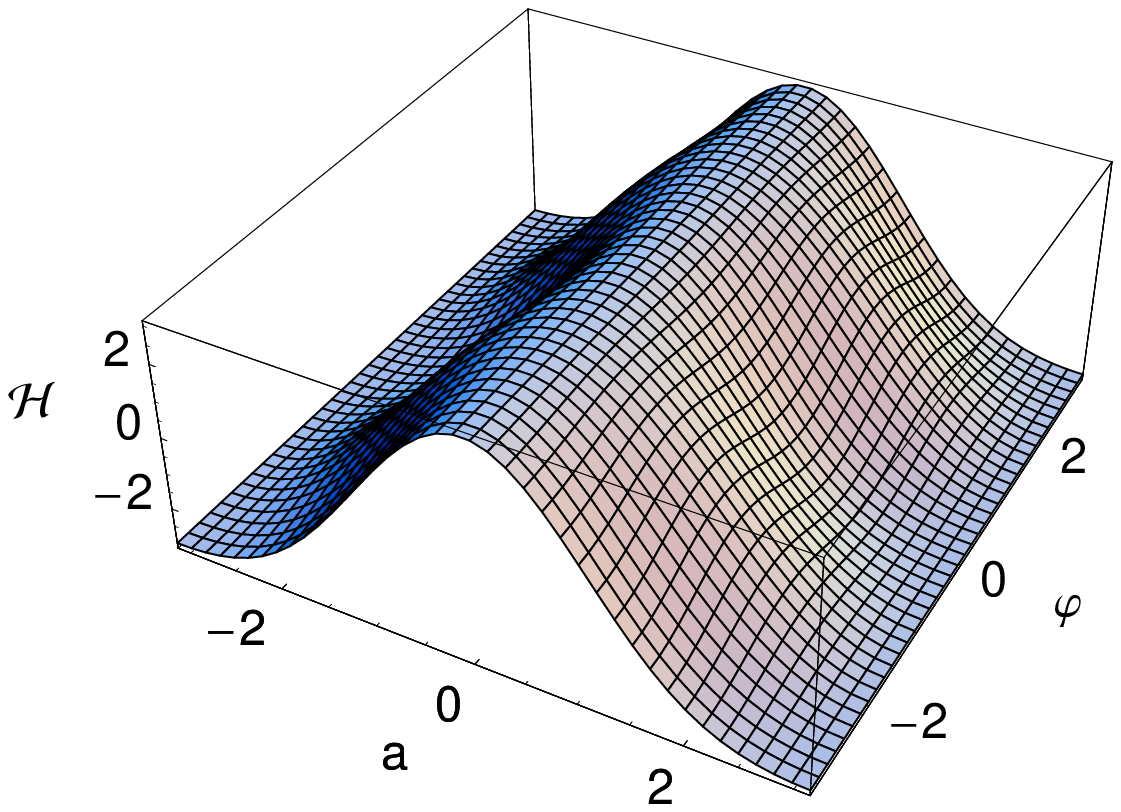}
\end{center}
\caption{Hamiltonian function $H$ in Eq. (\ref{e19}), showing distinct
qualitative properties for different values of the parameter $\mu$: $\mu
=\sqrt{0.5}$ (upper panel), $\mu=\sqrt{3}$ (lower panel).}%
\end{figure}In order to access the low-dimensional dynamical behavior
derivable from Eqs. (\ref{e1}--\ref{e2}), we follow Refs. \cite{Misra1} and
\cite{Batra} considering a few-modes expansion through the ansatz
\begin{align}
E  &  =\sqrt{N}\sin\left(  \frac{a}{2}\right)  \exp(i\theta_{0})+\sqrt{2N}%
\cos\left(  \frac{a}{2}\right)  \cos(kx)\exp(i\theta_{1})\,,\label{eq3}\\
n  &  =N+n_{1}\cos(kx)\,, \label{e4}%
\end{align}
where $N$ is the conserved plasmon number, $k$ is the fixed wave-number of the
excitation and $a=a(t)$, $\theta_{0}=\theta_{0}(t)$, $\theta_{1}=\theta
_{1}(t)$ and $n_{1}=n_{1}(t)$ are real time-dependent quantities. It is
important to notice that Eq. (\ref{eq3}) implies $|E|^{2}=N+$oscillatory
terms, in agreement with the conservation of plasmon number.

Inserting Eqs. (\ref{eq3}--\ref{e4}) into Eqs. (\ref{e1}--\ref{e2}) and
computing the zero wave-number term and the coefficient of the term
proportional to $\cos(kx)$ yields four equations, corresponding to the real
and imaginary parts. Further, the higher-order harmonics proportional to
$\cos(2kx)$ are disregarded, in the spirit of the present few modes
approximation. The real and the imaginary parts of the independent term gives
\begin{align}
\dot{a}  &  =-\sqrt{2}\,n_{1}\sin\varphi\,,\label{e5}\\
\dot{\theta_{0}}  &  =-N-\frac{\sqrt{2}}{2}\,n_{1}\cos\varphi\,\cot\left(
\frac{a}{2}\right)  \,, \label{e6}%
\end{align}
where the `dot' represents derivative with respect to time $t$ and
\begin{equation}
\varphi=\theta_{0}-\theta_{1}\,. \label{e7}%
\end{equation}
Now, considering the real part of the coefficient of the term proportional to
$\cos(kx)$ we get
\begin{equation}
\dot{\theta_{1}}=-N-k^{2}(1+H^{2}k^{2})-\frac{\sqrt{2}}{2}\,n_{1}\cos
\varphi\,\tan\left(  \frac{a}{2}\right)  \,. \label{e8}%
\end{equation}
The imaginary part of the coefficient or the term proportional to $\cos(kx)$
provides no further information. The combination of Eqs. (\ref{e6}--\ref{e8})
gives
\begin{equation}
\dot{\varphi}=k^{2}(1+H^{2}k^{2})-\sqrt{2}\,n_{1}\cos\varphi\,\cot a\,.
\label{e9}%
\end{equation}
To close the system, it is necessary to substitute the ansatz of Eqs.
(\ref{eq3}--\ref{e4}) also into Eq. (\ref{e2}). Ignoring again higher-order
harmonics yields
\begin{equation}
\ddot{n}_{1}+k^{2}(1+H^{2}k^{2})\,n_{1}=-\sqrt{2}\,Nk^{2}\cos\varphi\,\sin
a\,. \label{e10}%
\end{equation}
The system of Eqs. (\ref{e5}),(\ref{e9}) and (\ref{e10}) provides a
finite-dimensional closed set of ordinary differential equations for $a$,
$\varphi$ and $n_{1}$, whose pro\-per\-ties will be analyzed numerically in
the next Section.

Of particular interest is the case where the density fluctuations respond
adiabatically to excitations. In this si\-tua\-tion the second derivative term
in Eq. (\ref{e10}) can be neglected, yielding
\begin{equation}
n_{1}=-\frac{\sqrt{2}\,N}{1+H^{2}k^{2}}\,\cos\varphi\,\sin a\,. \label{e11}%
\end{equation}
Inserting this last result into Eqs. (\ref{e5}) and (\ref{e9}), we get the
two-dimensional dynamical system
\begin{align}
\dot{a}  &  =\frac{N}{1+H^{2}k^{2}}\,\sin2\varphi\,\sin a\,,\label{e12}\\
\dot{\varphi}  &  =k^{2}(1+H^{2}k^{2})+\frac{2N}{1+H^{2}k^{2}}\,\cos
^{2}\varphi\,\cos a\,. \label{e13}%
\end{align}
Actually, Eqs. (\ref{e12}) and (\ref{e13}) contain only one relevant free
parameter $\mu$, as can be proved by introducing the rescaled time variable
\begin{equation}
\tau=\frac{N\,t}{1+H^{2}\,k^{2}}\,, \label{e14}%
\end{equation}
so that
\begin{align}
da/d\tau &  =\sin2\varphi\,\sin a\,,\label{e15}\\
d\varphi/d\tau &  =\mu^{2}+2\cos^{2}\varphi\,\cos a\,, \label{e16}%
\end{align}
where
\begin{equation}
\mu=\frac{k\,(1+H^{2}k^{2})}{\sqrt{N}}\,. \label{e17}%
\end{equation}
In this formulation, the role of quantum effects, contained in the free
parameter $H$, are hidden in the rescaled time $\tau$ and also in $\mu$. Thus,
the quantum effects tend to slow down the dynamics, since a larger $H$ causes
$d\tau/dt$ to become smaller in magnitude. Moreover, it follows from Eq.
(\ref{e16}) that\ the fixed points can exist only when $\mu\leq\sqrt{2}$. But,
since $\mu$ is a monotonically increasing function of $H$, the conclusion is
that quantum effects tend to suppress the existence of equilibrium points.

In addition, Eqs. (\ref{e15}-\ref{e16}) form a completely integrable dynamical
system, as can be better seen in writing it in the generalized Hamiltonian
form
\begin{equation}
\frac{da}{d\tau}=J\,\frac{\partial\mathcal{H}}{\partial\varphi}\,,\qquad
\frac{d\varphi}{d\tau}=-J\,\frac{\partial\mathcal{H}}{\partial a}\,,
\label{e18}%
\end{equation}
where $J=\csc a$ and the first integral $H$ playing the role of Hamiltonian
function is
\begin{equation}
\mathcal{H}=\mu^{2}\cos a-\cos^{2}\varphi\,\sin^{2}a\,. \label{e19}%
\end{equation}
The system (\ref{e18}) can be written in canonical Hamiltonian form using
Darboux coordinates $(u,v)=(-\cos a,\,\varphi)$, so that $du/d\tau=\partial
H/\partial v,dv/d\tau=-\partial H/\partial u$. Most importantly, Eq.
(\ref{e18}) is manifestly non-chaotic. Notice, however, that the level
surfaces of the constant of motion $H$ are not compact, and thereby violating
a necessary condition for Liouville integrability. In addition, the Poisson
structure is singular in the sense that $J$ is not well-defined for $a=l\pi$,
where $l$ is an integer. Figure 1 exhibits some typical graphs of $H$ for
$\mu=\sqrt{0.5}$ and $\mu=\sqrt{3}$ showing distinct qualitative properties of
the constant of motion.

On the other hand, the adiabatic limits of the simplified finite-dimensional
system and the original infinite-dimensional system are to be compared. For
the infinite-dimensional system, the adiabatic limit is get disregarding the
second-order time derivative of the density fluctuation in Eq. (\ref{e2}). The
resulting equations are
\begin{align}
i\frac{\partial E}{\partial t}+\frac{\partial^{2}E}{\partial x^{2}}+|E|^{2}E
&  =H^{2}\left(  \frac{\partial^{4}E}{\partial x^{4}}+E\frac{\partial^{2}%
n}{\partial x^{2}}\right)  \,,\label{e20}\\
H^{2}\frac{\partial^{2}n}{\partial x^{2}}-n  &  =|E|^{2}\,, \label{e21}%
\end{align}
which in the formal classical limit ($H\rightarrow0)$ reduce to the usual
NLSE, which is well-known to be completely integrable. However, in the quantum
case, the adiabatic limit still shows a coupled nonlinear system, whose
properties are not yet completely known. In contrast, the simplified dynamics
associated to the present few-modes ansatz has shown to be integrable. Notice
that the system (\ref{e20}--\ref{e21}) can be decoupled taking both the
semiclassical $(H\ll1)$ as well as adiabatic limits, so that the substitution
$n=-|E|^{2}$ is allowed in the right-hand side of Eq. (\ref{e20}). In this
case one obtains
\begin{equation}
i\frac{\partial E}{\partial t}+\frac{\partial^{2}E}{\partial x^{2}}%
+|E|^{2}E=H^{2}\left(  \frac{\partial^{4}E}{\partial x^{4}}-E\,\frac
{\partial^{2}|E|^{2}}{\partial x^{2}}\right)  . \label{e22}%
\end{equation}
Equation (\ref{e22}) can be used as the starting point for studying quantum
perturbations of the classical NLS soliton solutions \cite{Misra2}.

\section{Numerical results}

The system of Eqs. (\ref{e5}), (\ref{e9}) and (\ref{e10}) can be recast as
\begin{align}
\dot{x}_{1}  &  =-\sqrt{2}\,x_{3}\sin x_{2}\,,\label{e23}\\
\dot{x}_{2}  &  =k^{2}\,(1+H^{2}k^{2})-\sqrt{2}\,x_{3}\cos x_{2}\,\cot
x_{1}\,,\label{e24}\\
\dot{x}_{3}  &  =x_{4}\,,\label{e25}\\
\dot{x}_{4}  &  =-k^{2}\,(1+H^{2}k^{2})\,x_{3}-\sqrt{2}\,Nk^{2}\cos
x_{2}\,\sin\,x_{1}\,, \label{e26}%
\end{align}
where, for convenience, we have redefined the variables as $a=x_{1}%
,\varphi=x_{2},n_{1}=x_{3},\dot{n}_{1}=x_{4}$.

Eqs.(\ref{e23}--\ref{e26}) were numerically solved by using the sixth order
Runge-Kutta-Fehlberg scheme with the step length $h=0.01$ and initial values
as $x_{1}=0.1$, $x_{2}=0.2$, $x_{3}=0.3$, $x_{4}=0.1$. A Monte-Carlo search on
the parameter space was conducted to find the possible dynamics of the system,
namely periodic, chaotic or hyperchaotic regimes. \begin{figure}[ptb]
\begin{center}
\includegraphics[height=2.5in,width=2.5in]{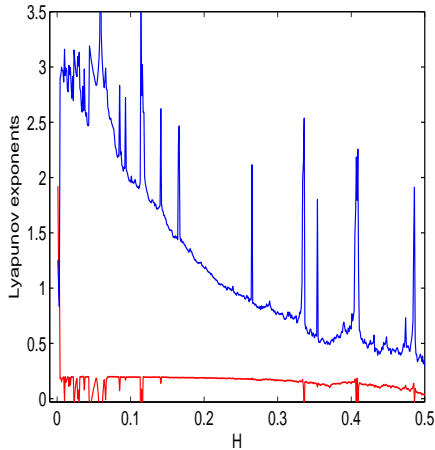}
\end{center}
\caption{Two largest Lyapunov exponents with respect to $H$ and for constant
$N=1.5$ and $k=0.8.$}%
\end{figure}\begin{figure}[ptb]
\begin{center}
\includegraphics[height=2.5in,width=2.5in]{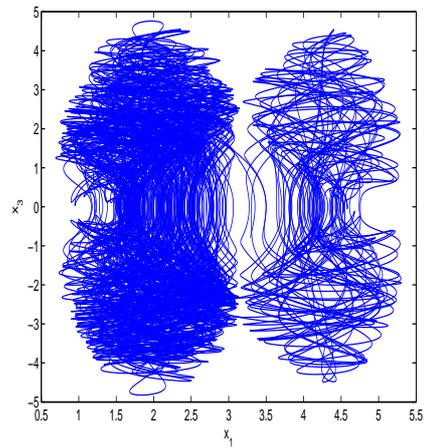}
\end{center}
\caption{Hyperchaotic phase portrait of the system (\ref{e23})-(\ref{e26}) in
the $x_{1}-x_{3}$ plane for $H=0.2,N=1.5$ and $k=0.8$.}%
\end{figure}In order to establish the existence of irregular dynamics we have
investigated the Lyapunov exponents spectra \cite{Wolf}. Chaos or hyperchaos
is characterized by the presence of one or two positive largest Lyapunov
exponents respectively. It is sufficient to calculate only the three largest
Lyapunov exponents.\ \ These largest Lyapunov exponents were calculated by
integrating Eqs. (\ref{e23}-\ref{e26}) in order to\ have average estimates of
them over the attractors \cite{NumericalComputation}. Figure 2 shows examples
of two largest Lyapunov exponents with respect to the quantum parameter $H$
and for constant $N=1.5$ and $k=0.8$ such that $\mu\leq\sqrt{2}$ is satisfied.
We observe that the system is chaotic (one positive exponent) for a small
range of $H$, while it is hyperchaotic (two positive exponents) for a wide
range of values of $H$.\ In the classical case, i.e., for $H=0$, we also find
two positive Lyapunov exponents indicating the hyperchaotic features even in
the classical ZEs \cite{Zakharov} as well. It is to be noted that the system
may experience some other features at the points where some peaks of the first
(maximum) Lyapunov exponent seem to be correlated with the depletions of the
second exponent. However, since the variations of the exponents are shown with
respect to a system parameter, namely $H$, there may be a possibility for the
existence of chaos rather than the hyperchaotic orbits at those points,
especially where the second Lyapunov has negative peaks. Figure 3 shows an
example of the hyperchaotic dynamics of the system in the $x_{1}-x_{3}$ plane
for $H=0.2,N=1.5$ and $k=0.8.$ It seems to show an intermittent switching
between two regions of the phase space, one to the left and other to the right
of $x_{1}\approx3.2$. These can be verified from the corresponding time series
for $x_{1}$ and $x_{3}$ as shown in Fig. 4. Here the upper panel clearly
explains this intermittency near $x_{1}=3.2$, and both the time series show
the aperiodic nature of the system which are very common in the context of
chaotic dynamics. The chaotic dynamics of the system can be established by calculating the power
spectrum corresponding to the variable $x_{3}.$ Here one can measure the power
spectrum as the square of the modulus of the complex Fourier coefficients
corresponding to $x_{3}$ with frequency as the inverse of the period of the
signal \cite{Pikovskii}$.$ However, measuring the amplitude of the pulse gives
rise a slight modification of the power of the same obtained\ by using the
Fast Fourier Transform (FFT). The latter is used to compute the discrete
Fourier Transform (DFT) of the signal, as well as the magnitude and phase of
the transformed signal \cite{MagnitudePhase}. The `abs' function is used to
obtain the magnitude of the data and the `angle' function to obtain the phase
information \cite{Kodba}, and unwrap in order to remove the phase jumps
greater than $\pi$ to their $2\pi$ complement. These are illustrated in Fig.
5. This figure clearly shows the chaotic dynamics of the waveforms with
low-frequency and high amplitudes, as well as exhibits the broadband spectral
features of the system.
\begin{figure}[ptb]
\begin{center}
\includegraphics[height=3.0in,width=4.0in]{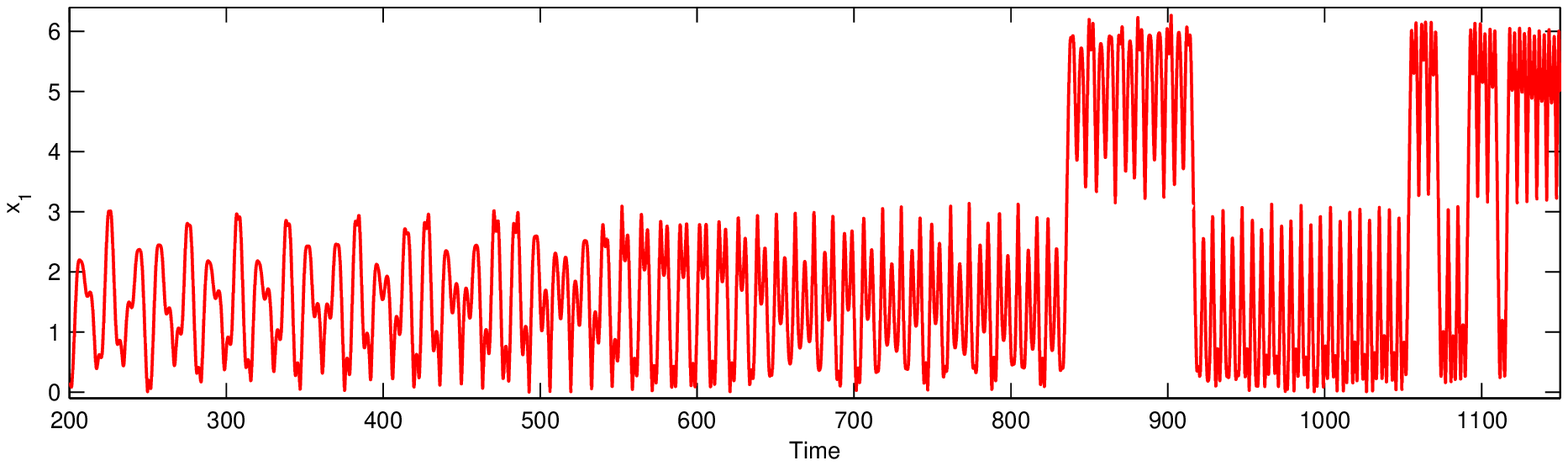}\newline%
\includegraphics[height=3.0in,width=4.0in]{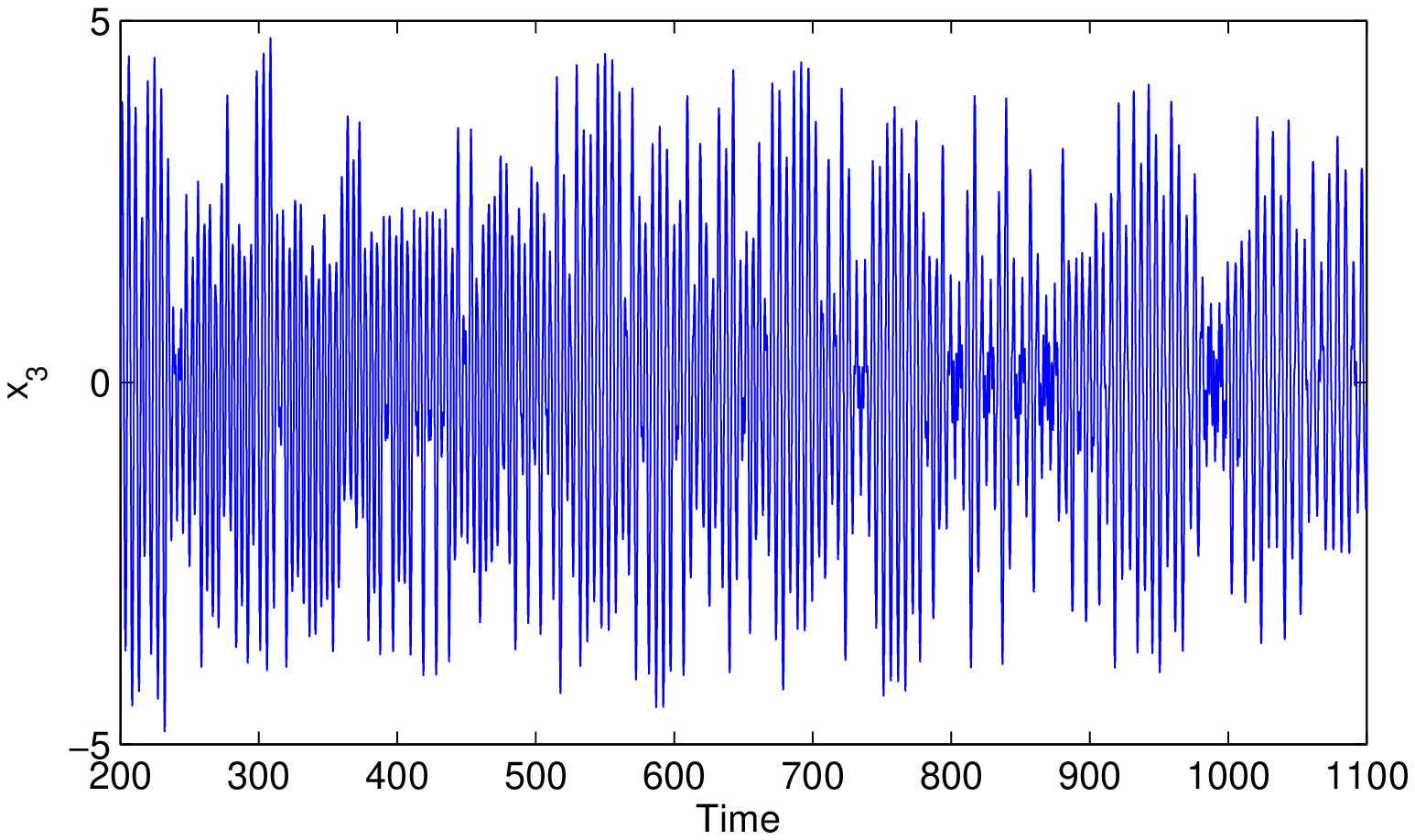}
\end{center}
\caption{The time series corresponding to the variables $x_{1}$ (upper panel)
and $x_{3}$ (lower panel). The parameter values are the same as in Fig. 3.}%
\end{figure}
The route to chaos from a single periodic orbit can be explained
by means of a one-parameter bifurcation analysis in the domains of dynamical
variables. We have considered one fixed value of $H=0.2,$ two different values
of $k,$ namely $k=0.9,0.95$ and allowed the plasmon number $N$ to vary so as
to satisfy the inequality $\mu\leq\sqrt{2}$. The three-dimensional pictorial
views of the bifurcation diagram with $x_{1},$ $x_{3}$ and $N$ are shown in
Fig. 6 for distinct values of $k$. These explain that the system loses its
stability through a supercritical Hopf-bifurcation (hb), which gives the birth
to periodic limit cycles at $N\approx0.78$ for $H=0.2,k=0.9$ (upper panel)
and\ $N\approx0.6$ for $H=0.2,k=0.95$ (lower panel). 
 \begin{figure}[ptb]
\begin{center}
\includegraphics[height=3.0in,width=4.0in]{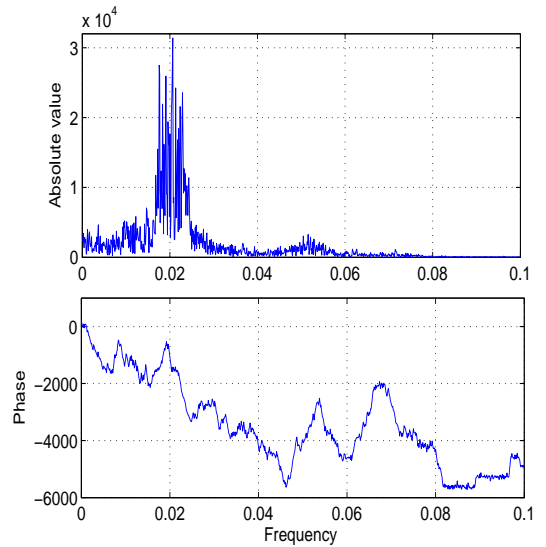}
\end{center}
\caption{The amplitude (absolute value, see the upper panel) and the phase
(lower panel) of the hyperchaotic signal corresponding to the variable $x_{3}%
$. The parameter values are the same as in Fig. 3.}%
\end{figure}\begin{figure}[ptb]
\begin{center}
\includegraphics[height=4in,width=4in]{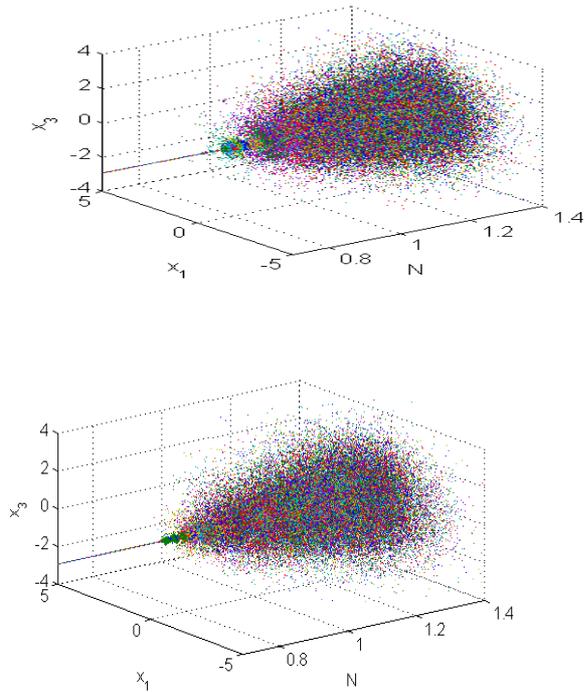}
\end{center}
\caption{The snap shots of an animated movie showing the three-dimensional
views of the bifurcation diagrams with variation of the parameter $N$ and for
a fixed $H=0.2.$ Two different values of $k$ are$:$ $k=0.9$ (upper panel) and
$k=0.95$ (lower panel). Different colors may correspond to the pixels of
different projections on the \textquotedblleft floor\textquotedblright.}%
\end{figure}
The nature of the supercritical Hopf-bifurcation can be seen in the $x_{1}$-$x_{2}$ plane from
Fig. 7 for $N=1.5,$ $H=0.2$ and different $k$: (i) $k=0.3$ (upper panel) and
(ii) $k=0.5$ (lower panel). Notice that in the case of Hopf-bifurcation, we
have generated a program with varying $k$ from $k=0.1$ to $1.0$. Figure 7
shows illustrations of two different values of $k.$ The trajectories in both the upper
and lower panel of Fig. 7 are generated with different initial conditions, and finally
they converge to attracting or repelling limit cycles. In both the cases the initial values
 are taken from the unstable sets. For values of $k$ larger
than $k=0.5$ no more attracting or repelling limit cycle is found to coexist.
From Fig. 7 one can also observe that a limit cycle (the blue curves or e.g.,
the orbits appeared first from the bottom at $x_{\mathbf{1}}<0$ and at
$x_{\mathbf{1}}>0$ ) is surrounded by the unstable equilibriums (red colored
curves). To further examine the chaotic features, one can also calculate the
Fourier transform of the signal \cite{Pikovskii} corresponding to the variable
$x_{1}$. Figure 8 represents the distribution of the Fourier coefficients in
the complex plane corresponding to the chaotic signal $x_{1}$for $H=0.2,N=1.5$
and $k=0.8$. It shows how wide and dense are the Fourier coefficients in the
complex plane. This distribution could be useful to quantify chaos from
numerical points of view. However, the detail analysis is beyond the scope of
the present work.
\begin{figure}[ptb]
\begin{center}
\includegraphics[height=2.0in,width=3in]{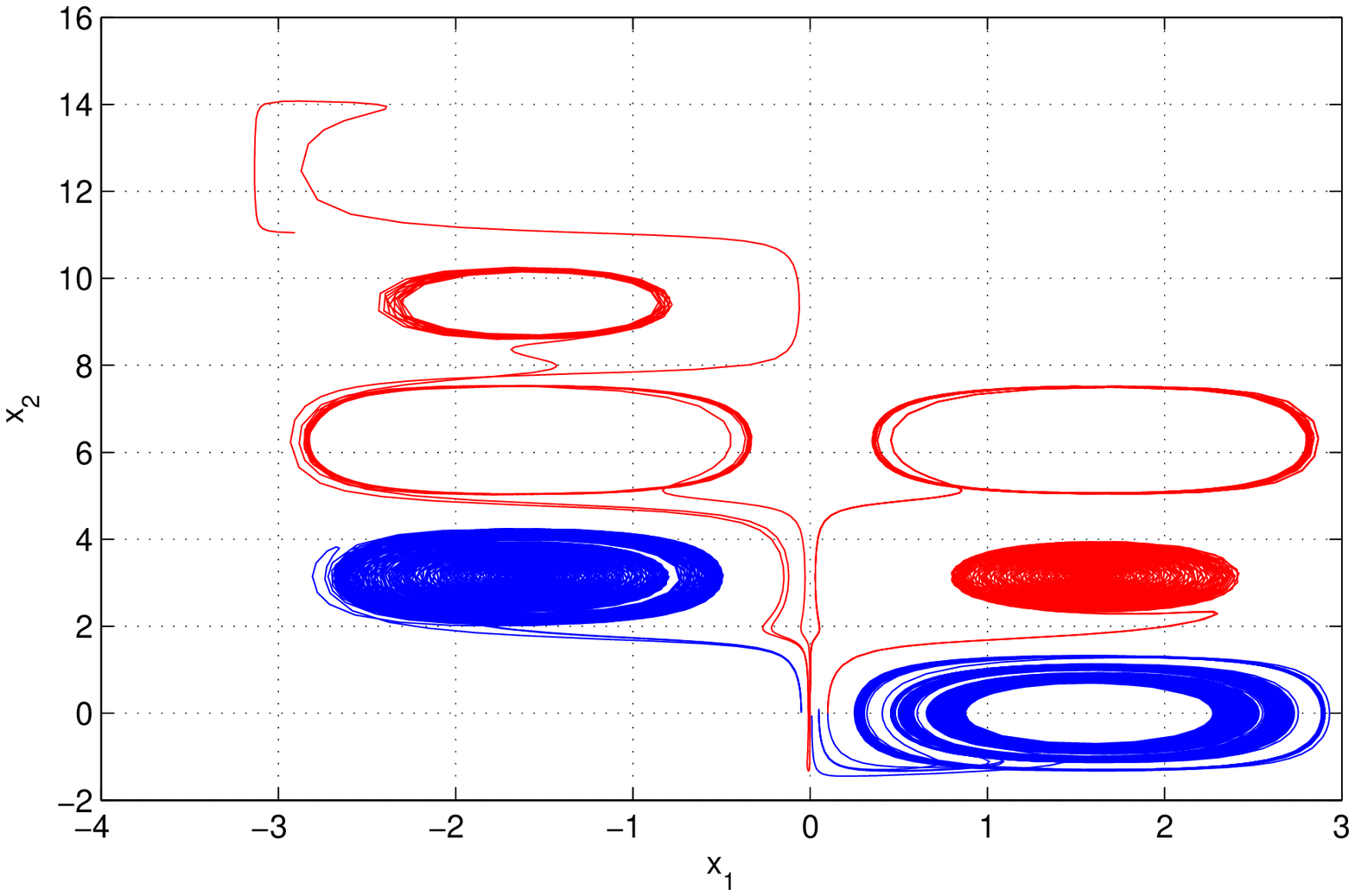}\newline%
\includegraphics[height=2.0in,width=3in]{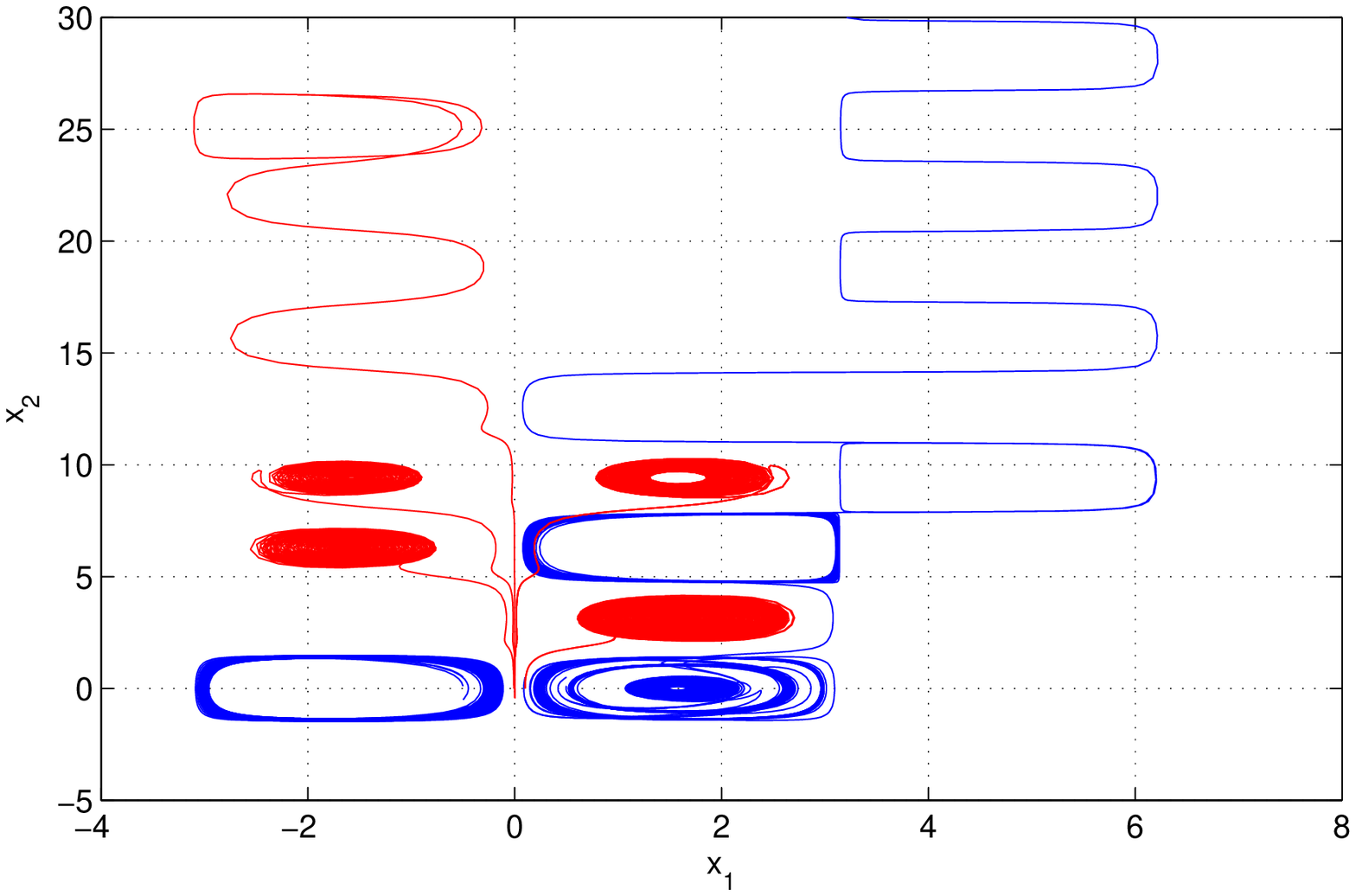}
\end{center}
\caption{Supercritical Hopf-bifurcation in the $x_{1}-x_{2}$ plane for
$N=1.5,H=0.2$ and different $k$: (i) $k=0.3$ (upper panel) (ii) $k=0.5$ (lower
panel).}%
\end{figure}
\begin{figure}[ptb]
\begin{center}
\includegraphics[height=2.5in,width=2.5in]{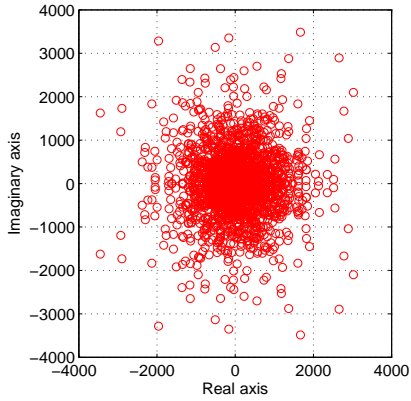}
\end{center}
\caption{Distribution of the Fourier coefficients in the complex plane
corresponding to the chaotic signal $x_{1}$ for $H=0.2,N=1.5$ and $k=0.8$.}%
\end{figure}

\section{Conclusion}

The nonlinear interaction of quantum Langmuir waves and quantum ion-acoustic
waves is analyzed in terms of a superposition of three interacting wave modes
in Fourier space. Previous works on both classical \cite{Sharma,Batra} and
quantum \cite{Misra1} Zakharov equations have been rectified and modified. The
chaotic behaviors of the reduced temporal system have been identified by the
analysis of Lyapunov exponent spectra as well as by the analysis of power
spectrum. The hyperchaos has been characterized by the presence of two
positive Lyapunov exponents. Also, the route to chaos from a single periodic
orbit is analyzed by means of one-parameter bifurcation analysis. Moreover,
the reduced temporal dynamics is shown to be integrable in the adiabatic
limit. This is in contrast to the adiabatic limit of the spatiotemporal
dynamics described by the QZEs whose integrable properties deserves further
study; whereas the finite-dimensional reduced system is shown to evolve into
periodic, chaotic as well as hyperchaotic orbits \cite{Misra1}. To conclude,
we believe that the results presented in this work would be helpful for better
understanding the salient features of such 1D QZEs.\newline\newline
ACKNOWLEDGMENTS\newline\newline A. P. M. is grateful to the Kempe Foundation,
Sweden for support. F.H. acknowledges support from Ume\aa \ University and the
Kempe Foundation.

\ \ \ \


\begin{thebibliography}{99}
\bibitem {MagnitudePhase}The MATLAB codes used to compute the amplitude and
the phase are as follows:

h=fft(g); m=abs(h); p=unwrap(angle(h)); \% Plot the magnitude and phase.

f=(0:length(g)-1)'*100/length(h); subplot(2,1,1), plot(f,m); ylabel('Abs.
Magnitude'), grid on;

subplot(2,1,2), plot(f,p*180/pi); ylabel('Phase [Degrees]'),grid on;
xlabel('Frequency [Hertz]');
\end{thebibliography}
\end{document}